\documentclass[12pt]{article}
\usepackage{epsfig}

\def\beq{\begin{equation}}
\def\eeq#1{\label{#1}\end{equation}}
\def\eeqn{\end{equation}}

\def\beqa{\begin{eqnarray}}
\def\eeqa#1{\label{#1}\end{eqnarray}}
\def\eeqan{\end{eqnarray}}

\let\bar=\overbar

\def\O{{\cal O}}

\def\Dslash{\not{\hbox{\kern-4pt $D$}}}
\def\dslash{\not{\hbox{\kern-2pt $\del$}}}

\def\msb{{\bar{\ssstyle M \kern -1pt S}}}

\def\Title#1{\begin{center} {\Large {\bf #1} } \end{center}}

\begin{document}

\Title{A Colored Zoo of \\Quasi-particles and Light Glueballs}

\bigskip\bigskip


\begin{raggedright}

{\it Francesco Sannino\index{Sannino, F.}\\ {\rm NORDITA},\\
Blegdamsvej 17,  DK-2100 \\ Copenaghen \O, DENMARK}
\bigskip\bigskip
\end{raggedright}

\section{Electroweak for CFL}

\label{3f}

Here I present a compendium of the effective Lagrangians for three
and two flavors QCD describing the relevant color superconductive
light degrees of freedom and their electroweak interactions.
However I will not consider some more recent developments
involving new phases which are presented by other participants in
this conference. Possible phenomenological applications include
the description of quark stars, neutron star interiors, the
physics near the core of collapsing stars and supernova explosions
\cite{RW,Hong:2001gt,OS}.

At high density, in the 3 flavor case, the symmetry group
$SU_{L}(3)\times SU_{R}(3)\times SU_{c}(3)$ breaks spontaneously
to $SU_{c+L+R}(3)$ leaving behind $16$ Goldston bosons. However,
being $SU_c(3)$ a gauge group 8 Goldstone bosons are absorbed in
the longitudinal components of the massive gluons. So we are left
with 8, not colored, physical massless Goldstone bosons. They can
be encoded in the unitary matrix \cite{CG} transforming linearly
under the left-right flavor rotations $\displaystyle{U\rightarrow
g_{L}Ug_{R}^{\dagger}}$ with $g_{L/R}\in SU_{L/R}(N_{f})$.  $U$
satisfies the non linear realization constraint $UU^{\dagger }=1$.
We also require ${\rm det}U=1$. In this way we avoid discussing
the axial $U_{A}(1)$ anomaly at the effective Lagrangian level.
(See Ref.~\cite{SS} for a general discussion of trace and
$U_{A}(1)$ anomaly). We have
$\displaystyle{U=\exp\left[i\Phi/F\right]}$ with $\Phi
=\sqrt{2}\Phi ^{a}t^{a}$ representing the $8$ Goldstone bosons and
$\displaystyle{{\rm Tr}\left[ t^{a}t^{b}\right] =\delta ^{ab}/2}$.
$F$ is the Goldstone bosons decay constant at finite density.

The effective Lagrangian, for massless quarks, globally invariant
under chiral rotations (up to two derivatives) and invariant only
under the rotational subgroup of the Lorentz group is:
\begin{equation}
L=\frac{F^{2}}{2}\,{\rm Tr}\left[\dot{U}\dot{U}^{\dagger} - v^2
{\vec{\nabla} U \cdot \vec{\nabla}U^{\dagger}}\right]\ .
\end{equation}
Clearly by rescaling the vector coordinates $\vec{x}\rightarrow
\vec{x}/v$ we can recast the previous Lagrangian in the form
$\displaystyle{ L=\frac{F^{2}}{2}\,{\rm Tr}\left[ \partial _{\mu
}U\partial ^{\mu }U^{\dagger }\right] \label{ef}}$.

 We now augment
the theory by the time-honored Wess-Zumino term which is compactly
written using the language of differential forms. It is useful to
introduce the algebra valued Maurer-Cartan one forms:
\begin{equation}
\alpha =\left( \partial _{\mu }U\right) U^{-1}\,dx^{\mu }\equiv
\left( dU\right) U^{-1}\ , \qquad \quad
\beta=U^{-1}dU=U^{-1}\alpha U \ ,  \label{MC}
\end{equation}
which transform, respectively, under the left and right $SU(N_f)$
flavor group.
 The
Wess-Zumino effective action is
\begin{equation}
\Gamma _{WZ}\left[ U\right] =C\,\int_{M^{5}}{\rm Tr}\left[ \alpha
^{5}\right] \ , \qquad   C=-i\frac{N_{c}}{240\pi ^{2}}\
.\label{WZ}
\end{equation}
 $C$ assumes the same value then the one used at zero density and $N_{c}$ is
the number of colors (fixed to be 3 in this case). The coefficient
is fixed by saturating the global anomalies at the effective
Lagrangian level \cite{CDS2001}. The 3 flavor case respects the 't
Hooft global anomaly matching conditions at non zero density. A
relevant feature is that $\alpha$, being a differential form, is
unaffected by coordinate rescaling (actually topological terms
being independent on the metric are unaffected by medium effects),
and hence the Wess-Zumino term is not modified at finite matter
density.

At this point it is important to note\footnote{I thank D.K.~Hong
for helping me clarifying this point.} that in reference
\cite{HRZ} the coefficient of the Wess-Zumino term actually agrees
with the one presented here \cite{S,CDS2001} while it is the one
presented/computed in \cite{RWZ} which disagrees by a factor
three. A relevant consequence of our results is associated to the
solitonic (Skyrme) sector of the effective Lagrangian theory. In
fact now we have the same winding number as for ordinary QCD and
hence we get massive excitations, which after collective
quantization, describe spin half particles with the same quantum
numbers of ordinary baryons.


 Next we
extend the previous effective Lagrangian by incorporating the
electroweak intermediate vector mesons as external fields.
\begin{eqnarray}
DU&=&\partial U-i\frac{g}{\sqrt{2}}\left[ W^{+}\tau ^{+}+W^{-}\tau
^{-}\right] U-i\frac{g}{\cos \theta _{W}}\,Z^{0}\,\left[ \tau
^{3}U-\sin ^{2}\theta _{W} \left[ Q,U\right] \right]  \nonumber \\
&&-i\widetilde{e}\,\widetilde{A}\,\left[ Q,U\right]
-i\widetilde{e}\,\tan \theta \widetilde{G}^{8}\,\left[ Q,U\right]
\label{elw}
\end{eqnarray}
where $\theta _{W}$ is the electroweak angle and
\begin{equation}
\tau ^{+}=\left(
\begin{array}{ccc}
0 & 1 & 0 \\
0 & 0 & 0 \\
0 & 0 & 0
\end{array}
\right) \ ,\qquad \tau ^{-}=\left(
\begin{array}{ccc}
0 & 0 & 0 \\
1 & 0 & 0 \\
0 & 0 & 0
\end{array}
\right) \ ,\qquad \tau ^{3}=\frac{1}{2}\left(
\begin{array}{ccc}
1 & 0 & 0 \\
0 & -1 & 0 \\
0 & 0 & -1
\end{array}
\right) \ . \label{tau}
\end{equation}
The last two terms in Eq.~(\ref{elw}) describe the, non anomalous,
interaction of the Goldstone bosons with respectively the physical
massless photon and the physical massive eight gluon. The physical
photon and eight gluon states are related to the, in vacuum, ones
via:
\begin{eqnarray}
\widetilde{G}^{8} =\cos \theta G^{8}+\sin \theta A\ , \qquad
\widetilde{A} =-\sin \theta G^{8}+\cos \theta A\ ,
\end{eqnarray}
with $\cos \theta =\sqrt{3}{g_{s}}/{\sqrt{3g_{s}^{2}+4e^{2}}}$.
Clearly $\widetilde{e}=e\,\cos \theta $ is the finite density new
electric charge.

The $\pi^0\rightarrow \gamma \gamma$ Lagrangian term is obtained
by gauging the Wess-Zumino action \cite{CDS2001} and is given by:
\begin{equation}
{\cal L}_{\pi^0\rightarrow\gamma\gamma}=-\frac{30}{4} e^2\, C\,i
{\rm Tr}\left[t^3 Q^2\right] \pi^0 \frac{\sqrt{2}}{F}
\epsilon^{\mu\nu\rho\sigma}F_{\mu\nu}F_{\rho\sigma} \ ,
\end{equation}
and $\pi^0=\Phi^3$. After substituting the, in vacuum, photon
field with its expression as function of the physical photon and
gluon \cite{CDS2001} we discover that the process is identical to
the in the vacuum one when replacing the, in vacuum, electric
charge with $\widetilde{e}$. It is also worth noticing that since
the mass of the physical eight gluon is larger than the pion mass
the process $\pi^0\rightarrow \widetilde{G}^8 \widetilde{G}^8$ is
barred kinematically.

\section{Electroweak for 2SC}

Here I summarize the effective low energy Lagrangian for two
flavors (when a 2SC phase sets is) which contains all of the
relevant degrees of freedom.

In 2SC the diquark condensate leaves invariant the following
symmetry group:\begin{equation} \left[ SU_{c}(2)\right] \times
SU_{L}(2)\times SU_{R}(2)\times \widetilde{U}_{V}(1) \
,\end{equation} where $\left[ SU_{c}(2)\right] $ is the unbroken
part of the gauge group. The $\widetilde{U}_{V}(1)$ generator
$\widetilde{B}$ is the following linear combination of the
previous $U_{V}(1)$ generator $B=\frac{1}{3}{\rm diag}(1,1,1)$ and
the broken diagonal generator of the $SU_{c}(3)$ gauge group
$T^{8}=\frac{1}{2\sqrt{3}}\,{\rm diag}(1,1,-2)$:
$\widetilde{B}=B-\frac{2\sqrt{3}}{3}T^{8} \label{residue}$. The
quarks with color $1$ and $2$ are neutral under $\widetilde{B}$
and consequently the condensate too ($\widetilde{B}$ is
$\sqrt{2}\widetilde{S}$ of Ref.~\cite{CDS}). The superconductive
phase for $N_{f}=2$ possesses the same global symmetry group of
the confined Wigner-Weyl phase \cite{S}. In Reference \cite{S}, it
was shown that the low-energy spectrum, at finite density,
displays the correct quantum numbers to saturate the 't~Hooft
global anomalies. In Reference \cite{HSaS} it was then seen, by
using a variety of field theoretical tools, that global anomaly
matching conditions hold for any cold but dense gauge theory.

The lowest lying excitations are protected from acquiring a mass
by the aforementioned constraints and dominate the low-energy
physical processes. Here, see \cite{CDS}, the relevant coset space
is $G/H$ with $G=SU_{c}(3)\times U_{V}(1)$ and $H=SU_{c}(2)\times
\widetilde{U}_{V}(1)$ is parameterized by
\begin{equation}{\cal V}=\exp (i\xi ^{i}X^{i})\ ,\end{equation}
where $\{X^{i}\}$ $i=1,\cdots ,5$ belong to the coset space $G/H$
and are taken to be $X^{i}=T^{i+3}$ for $i=1,\cdots ,4$ while
$X^{5}=B+\frac{\sqrt{3}}{3}T^{8}={\rm
diag}(\frac{1}{2},\frac{1}{2},0) \label{broken}$. $T^{a}$ are the
standard generators of $SU(3)$. The coordinates
\begin{equation}
\xi ^{i}=\frac{\Pi ^{i}}{f}\quad i=1,2,3,4\ ,\qquad \xi
^{5}=\frac{\Pi ^{5}}{\widetilde{f}}\ ,
\end{equation}
via $\Pi $ describe the Goldstone bosons.

${\cal V}$ transforms non linearly \begin{equation}{\cal V}(\xi
)\rightarrow u_{V}\,g{\cal V}(\xi )h^{\dagger }(\xi
,g,u)h_{\widetilde{V}}^{\dagger }(\xi ,g,u)\ ,
\label{nl2}\end{equation} with $u_{V}\in U_{V}(1)$, $g\in
SU_{c}(3)$, $h(\xi ,g,u)\in SU_{c}(2)$ and $h_{\widetilde{V}}(\xi
,g,u)\in \widetilde{U}_{V}(1)$. It is, also, convenient to define:
\begin{equation}
\omega _{\mu }=i{\cal V}^{\dagger }D_{\mu }{\cal V}\quad {\rm
with}\quad D_{\mu }{\cal V}=(\partial _{\mu }-ig_{s}G_{\mu }){\cal
V}\ ,
\end{equation}
with gluon fields $G_{\mu }=G_{\mu }^{m}T^{m}$.
Following \cite{CDS} we decompose $\omega _{\mu }$ into
\begin{equation}
\omega _{\mu }^{\parallel }=2S^{a}{\rm Tr}\left[ S^{a}\omega _{\mu
}\right] \quad {\rm and}\quad \omega _{\mu }^{\perp }=2X^{i}{\rm
Tr}\left[ X^{i}\omega _{\mu }\right] \ ,
\end{equation}
where $S^{a}$ are the unbroken generators of $H$ with
$S^{1,2,3}=T^{1,2,3}$, $S^{4}=\widetilde{B}\,/\sqrt{2}$. Summation
over repeated indices is assumed.

To be able to include the in medium fermions in the picture we
define:
\begin{equation}
\widetilde{\psi}={\cal V}^{\dagger }\psi \ ,  \label{mq}
\end{equation}
transforming as $\widetilde{\psi}\rightarrow
h_{\widetilde{V}}(\xi,g,u)h(\xi ,g,u)\widetilde{ \psi}$ and $\psi$
possesses an ordinary quark transformations (as Dirac spinor).

The simplest non-linearly realized effective Lagrangian describing
in medium fermions, the five gluons and their self interactions,
up to two derivatives and quadratic in the fermion fields is:
\begin{eqnarray}
{\cal L}=~ &&f^{2}a_{1}{\rm Tr}\left[ \,\omega _{0}^{\perp }\omega
_{0}^{\perp }-{\alpha }_{1}\vec{\omega}^{\perp
}\vec{\omega}^{\perp }\, \right] + f^{2}a_{2}\left[ {\rm Tr}\left[
\,\omega _{0}^{\perp }\,\right] {\rm Tr}\left[ \,\omega
_{0}^{\perp }\,\right] -{\alpha }_{2}{\rm Tr}\left[
\,\vec{\omega}^{\perp }\,\right] {\rm Tr}\left[
\,\vec{\omega}^{\perp }\, \right] \right]  \nonumber \\ &+&
b_{1}\overline{\widetilde{\psi }}i\left[ \gamma ^{0}(\partial
_{0}-i\omega _{0}^{\parallel })+\beta _{1}\vec{\gamma}\cdot \left(
\vec{ \nabla}-i\vec{\omega}^{\parallel }\right) \right]
\widetilde{\psi } + b_{2} \overline{\widetilde{\psi }}\left[
\gamma ^{0}\omega _{0}^{\perp }+\beta _{2} \vec{\gamma}\cdot
\vec{\omega}^{\perp }\right] \widetilde{\psi } \nonumber \\
&+&m_{M}\overline{\widetilde{\psi }^{C}}\gamma
^{5}(iT^{2})\widetilde{\psi }+ {\rm h.c.}\ , \label{cadusa}
\end{eqnarray}
where $\widetilde{\psi }^{C}=i\gamma ^{2}\widetilde{\psi }^{\ast
}$, $i,j=1,2 $ are flavor indices and \begin{equation}
T^{2}=S^{2}=\frac{1}{2}\left(
\begin{array}{ll}
\sigma ^{2} & 0 \\ 0 & 0
\end{array}
\right)\ , \end{equation} $a_{1},~a_{2},~b_{1}$ and $b_{2}$ are
real coefficients while $m_{M}$ is complex. The breaking of
Lorentz invariance to the $O(3)$ subgroup, following \cite{CG},
has been taken into account by providing different coefficients to
the temporal and spatial indices of the Lagrangian, and it is
encoded in the coefficients $\alpha $s and $\beta $s.

To construct the low energy effective theory for the electroweak
sector we need to generalize the one form $\omega_{\mu}=i{\cal
V}^{\dagger}D_{\mu} {\cal V}$ by introducing the new covariant
derivative:
\begin{equation}
D_{\mu }{\cal V}=(\partial _{\mu }-ig_s G_{\mu }-ig^{\prime} Y\,
B^y_{\mu}) {\cal V}=(\partial _{\mu }-ig_s G_{\mu }-ig^{\prime}
\frac{B}{2}\, B^y_{\mu}) {\cal V} \ .  \label{newD}
\end{equation}
$B^y_{\mu}$ is the standard hypercharge gauge field and is a
linear combination of the electroweak eigenstates associated to
the photon field $A_{\mu }$ and the neutral massive vector boson
$Z_{\mu }^{0}$, i.e.:
\begin{equation}
B_{\mu }^{y}=\cos \theta _{W}A_{\mu }-\sin \theta _{W}Z_{\mu
}^{0}\ ,
\end{equation}
with $\theta _{W}$ the standard electroweak angle.

After diagonalizing the full quadratic terms the new massless
eigenstate is interpreted as the, in medium, photon and is a
linear combination of the in vacuum photon and eight gluon:
\begin{equation}
\widetilde{A}_{\mu }=\cos \theta _{Q}A_{\mu }-\sin \theta
_{Q}G_{\mu }^{8}\ , \label{ThePhoton}
\end{equation}
with $\cos \theta
_{Q}=\sqrt{3}\frac{g_{s}}{\sqrt{3g_{s}^{2}+e^{2}}}$. The massive
state orthogonal to $\widetilde{A}_{\mu}$ is
$\displaystyle{\widetilde{G}_{\mu }^{8}=\cos \theta _{Q}G_{\mu
}^{8}+\sin \theta _{Q}A_{\mu}}$ and further mixes with $Z^{0}$ of,
in vacuum, mass $m_Z$. The new eigenstates are:
\begin{eqnarray}
\hat{G}_{\mu }^{8} =\cos \theta _{M}\widetilde{G}_{\mu }^{8}+\sin
\theta _{M}Z_{\mu }^{0}\ ,  \qquad  \widetilde{Z}_{\mu }^{0}
&=&\cos \theta _{M}Z_{\mu }^{0}-\sin \theta _{M}
\widetilde{G}_{\mu }^{8},
\end{eqnarray}
with $\tan 2\theta _{M} \approx \frac{2f^{2}\,\tan \theta
_{W}}{9m_{Z}^{2}}\left( a_{1}+2a_{2}\right)
\sqrt{3g_{s}^{2}+e^{2}}$ and we considered the physical limit
$m_{Z}^{2}\gg f^{2}$. In the same limit, as expected,
$\widetilde{G}^{8}$ and $Z^{0}$ do not mix much and we can use
them as physical eigenstates.

Having identified the correct physical eigenvalues and
eigenvectors we now turn to the quark sector.
We have $Q={\tau ^{3}+\frac{B-L}{2}}$ with $\tau =\tau _{L}+\tau
_{R}$
and  $\displaystyle{\widetilde{e}=e\,\cos \theta _{Q}}$ is the new
electric charge while $\widetilde{Q}$ is the associated electric
charge operator associated with the field $\widetilde{A}_{\mu }$:
\begin{equation}
\widetilde{Q}={\tau ^{3}\times {\mbox{\bf
1}}+\frac{\tilde{B}-L}{2}}=Q\times {\mbox{\bf
1}}-\frac{1}{\sqrt{3}}{\mbox{\bf 1}\times }T^{8}\ .
\label{newcharge}
\end{equation}

By expanding around ${\cal V}=1$ the relevant interaction terms we
derive \cite{CDS2001} the following full modified quark coupling
to the neutral weak current:
\begin{equation}
b_{1}\frac{g}{\cos \theta _{W}}\,Z_{\mu }^{0}\bar{\psi}\gamma
^{\mu }\left[ T_{L}^{3}-\sin \theta
_{W}^{2}Q-\frac{b_{2}-b_{1}}{3\,b_{1}}\sin \theta
_{W}^{2}(B+Q-\widetilde{Q})\right] \psi \ , \label{z0}
\end{equation}
where we used Eq.~(\ref{newcharge}) and $X^{5}=B+Q-\widetilde{Q}$.
Hence we discover that the neutral weak current is directly
affected by finite density effects even when neglecting the small
physical mixing between the eighth gluon $\widetilde{G}^{8}$ and
$Z^{0}$. We also observe that the modified electroweak coupling
only emerges for quarks with color indices 1 and 2, since
$\displaystyle{X^{5}=\frac{1}{2}{\rm diag}(1,1,0)}$. We expect
relevant phenomenological consequences of our result for the
cooling history of compact objects \cite{Shapiro} (see
Ref.~\cite{CDS2001} for more details).

\section{The $SU_c(2)$ Glueball Effective Lagrangian}
\label{Glueball}

The $SU_c(2)$ gauge symmetry does not break spontaneously and
confines. If the new confining scale \cite{rischke2k} is lighter
than the superconductive quark-quark gap the associated confined
degrees of freedom (light glueballs) \cite{OSPLB} can play,
together with the massless quarks a relevant role for the physics
of Quark Stars featuring a 2SC superconductive surface layer
\cite{OS}.

Defining by $H$ the mass-scale dimension 4 composite field
describing, upon quantization, the scalar glueball
\cite{schechter} we can construct the simplest light glueball
action, in medium, for the SU(2) color theory:
\begin{eqnarray}
S_{G-ball}=\int
d^4x\left\{\frac{c}{2}\sqrt{b}\,H^{-\frac{3}{2}}\left[\partial^{0}
H
\partial^{0}H - v^2
\partial^iH
\partial^iH\right]   -\frac{b}{2}
H\log\left[\frac{H}{\hat{\Lambda}^4}\right] \right\} \ .
\label{G-ball}
\end{eqnarray}
This Lagrangian correctly encodes the underlying $SU_c(2)$ trace
anomaly \cite{OSPLB}. The glueballs move with the same velocity
$v$ as the underlying gluons in the 2SC color superconductor.
$\hat{\Lambda}$ is the intrinsic scale associated with the theory
and can be less than or of the order of few MeVs
\cite{rischke2k,OSPLB} while $c$ is a constant of order unity. The
glueballs are light (with respect to the gap) and barely interact
with the ungapped fermions. They are stable with respect to the
strong interactions unlike ordinary glueballs. We define the
mass-dimension one glueball field $h$ via $\displaystyle{H=\langle
H\rangle \exp\left[{h}/{F_h}\right]}$. By requiring a canonically
normalized kinetic term for $h$ one finds $\displaystyle{F_h^2={c}
\sqrt{b\langle H\rangle}}$, while the glueball mass term is
$\displaystyle{M^2_h={\sqrt{b}}\sqrt{\langle H\rangle
/{2c}}={\sqrt{b}} \hat{\Lambda}^2/{2c\sqrt{e}}}$, which is clearly
of the order of $\hat{\Lambda}$.

Once created, the light $SU_c(2)$ glueballs are stable against
strong interactions but not with respect to electromagnetic
processes. The following decay width, at non zero baryon density,
was obtained by saturating the electromagnetic trace anomaly in
\cite{OSPLB}:
\begin{eqnarray}
 \Gamma\left[h\rightarrow
\gamma\gamma\right] \approx 1.2\times 10^{-2} \cos\theta_{Q}^4
\left[\frac{M_h}{1~{\rm MeV}}\right]^5~{\rm eV} \ ,
\end{eqnarray}
where $\alpha=e^2/4\pi \simeq 1/137$. {}For illustration purposes
we consider a glueball mass of the order of $1$~MeV which leads to
a decay time $\tau\sim~5.5\times~10^{-14}s$. We used
$\cos\theta_Q\sim~1$ since $\theta_Q \sim 2.5^{\circ}$
\cite{OSPLB}.

The present light glueball analysis is limited to the zero
temperature and high matter density case. However it is
interesting to investigate the effects of a non zero temperature.

\bigskip
It is a pleasure for me to thank R. Casalbuoni, Z. Duan, S.
D.~Hsu, R.~Ouyed and M.~Schwetz for sharing part of the work on
which this talk is based. This work is supported by the
Marie-Curie fellowship under contract MCFI-2001-00181.

\def\Discussion{
\setlength{\parskip}{0.3cm}\setlength{\parindent}{0.0cm}
     \bigskip\bigskip      {\Large {\bf Discussion}} \bigskip}
\def\speaker#1{{\bf #1:}\ }
\def\endDiscussion{}

\end{document}